\documentclass[11pt]{article}
\usepackage{amssymb,cite}
\usepackage{graphicx}
\usepackage{subfigure}
\usepackage[subfigure]{ccaption}
\usepackage{latexsym}
\usepackage{epsfig}
\usepackage{rotating}
\usepackage{lscape}

\newcommand{\eq}{\begin{equation}}
\newcommand{\en}{\end{equation}}
\newcommand{\be}{\begin{equation}}
\newcommand{\ee}{\end{equation}}
\newcommand{\eqa}{\begin{eqnarray}}
\newcommand{\ena}{\end{eqnarray}}
\newcommand{\ba}{\begin{eqnarray}}
\newcommand{\ea}{\end{eqnarray}}

\newcommand{\ZZ}{\hbox{{\rm Z{\hbox to 3pt{\hss\rm Z}}}}}

\begin{document}
\begin{titlepage}
\vskip0.5cm
\begin{flushright}
DFTT 4/07\\
\end{flushright}
\vskip0.5cm
\begin{center}
{\Large\bf DrosOCB: a high resolution map of conserved non coding sequences in Drosophila}
\end{center}
\vskip1.3cm
\centerline{Loredana~Martignetti$^{1,2}$, Michele~Caselle$^{1,2}$, Bernard~Jacq$^3$ and
Carl~Herrmann$^3$}
 \vskip1.0cm
 \centerline{\sl  1. Dipartimento di Fisica
 Teorica dell'Universit\`a di Torino and I.N.F.N.,}
 \centerline{\sl Via Pietro~Giuria 1, I-10125 Torino, Italy}
\centerline{\sl 2. Centre on Complex Systems in Biology and Molecular Medicine,}
\centerline{\sl Via Accademia Albertina 13, I-10125 Torino, Italy} 
\centerline{\sl e--mail: \hskip 1cm (caselle)(martigne)@to.infn.it}
\vskip0.1 cm
\centerline{\sl  3. Institut de Biologie du Développement de Marseille-Luminy,}
\centerline{\sl UMR6216 CNRS, Case 907, Parc Scientifique et Technologique de Luminy,}
\centerline{\sl 13288 Marseille, Cedex 9, France }
\centerline{\sl e--mail: \hskip 1cm (jacq)(herrmann)@ibdm.univ-mrs.fr}
\vskip0.4 cm

\begin{abstract}
Comparative genomics methods are widely used to aid the functional annotation of non coding DNA regions. However, aligning non coding sequences requires new algorithms and strategies, in order to take into account extensive rearrangements and turnover during evolution. Here we present a novel large scale alignment strategy which aims at drawing a precise map of conserved non coding regions between genomes, even when these regions have undergone small scale rearrangments events and a certain degree of sequence variability.\\ 
We applied our alignment approach to obtain a genome--wide catalogue of conserved non coding blocks (CNBs) between Drosophila melanogaster and 11 other Drosophila species. Interestingly, we observe numerous small scale rearrangement events, such as local inversions, duplications and translocations, which are not observable in the whole genome alignments currently available.\\ 

The high rate of observed low scale reshuffling show that this database of CNBs can constitute the starting point for several investigations, related to the evolution of regulatory DNA in Drosophila and the in silico identification of unannotated functional elements.
\end{abstract}
\end{titlepage}

\section{Background}
The functional annotation of eukaryotic DNA sequences represents a great challenge in post--genomic biological research. The identification of functional non--coding elements, such as untranslated regions (UTRs), genes for non--protein-coding RNAs, and cis-regulatory elements, is extremely difficult, as the rules governing their structure and function are far from being well undertood.\\

A great aid to functional annotation of genome sequences is provided by comparative genomics methods which, since a few years, have been extended also to non coding DNA regions.
The basic assumption of comparative genomic approach is that common features of two organisms are encoded within the DNA that is conserved between the species, due to purifying selection during evolution. According to the same assumption, the DNA sequences controlling the expression of genes that are regulated similarly in two related species should also be selected during evolution.\\

However, comparison of non coding sequences requires new algorithms and strategies to take into account the different evolutionary mechanisms affecting regulatory sequences. Recent studies examining the evolution of cis--regulatory modules in Drosophila, reveals that regulatory sequences may frequently evolve through compensatory gain and loss events in transcription factors binding sites, that produces little functional change\cite{ludwig},\cite{eisen}.
Great plasticity in the arrangement of binding sites within cis--regulatory modules is another remarkable evolutionary feature revealed to occur in vertebrates \cite{stupka}.\\

Once complete genomes from different species are available, a global alignment procedure is suitable to find a map of colinear conserved segments between the input sequences, descarding alignments that overlap or cross over. Global alignment methods are widely used to identify highly similar regions in the sequences which appear in the same order and orientation. On the contrary, local alignment algorithms are generally very useful in finding similarity between regions that may be related but are inverted or rearranged with respect to each other.

Recently, the novel notion of glocal alignment, a sophisticated combination of global and local methods, has been introduced \cite{brudno}. This class of alignment algorithms create a map that transforms one sequence into the other while allowing for rearrangement events. This procedure, at the base of Shuffled-LAGAN algorithm \cite{brudno2}, is able to take into account large scale genomic rearrangments, but fails at lower scale.

Here, we present an novel large scale alignment strategy which aims at drawing a precise map of conserved non-coding regions between genomes, even when these regions have undergone small
scale rearrangement events. Our procedure is optimized to take into account the great plasticity
of non coding DNA, such as shuffling and sequence variability of binding sites within functional
modules, low scale translocations, inversions and duplications.
We used a ``gene-centric" approch, in that it starts with a list of
orthologous genes between two species, and applies a local alignment algorithm to the corresponding flanking intergenic regions and intronic regions of these orthologous pairs. Hence, it is a
local alignment strategy but applied systematically on a genome-wide scale and, for this reason, we decided to  call it ``lobal". \\

The recent availability of 12 Drosophila species sequences and annotations \cite{aaa} offers a complete and reliable genomic dataset for developing and testing methods for comparative genomics of non coding DNA.
We applied our lobal alignment approach to align Drosophila melanogaster to several
other drosophila species (D. yakuba, D. pseudoobscura, D. virilis, ...), for which a reliable genome
build and annotation is available. 

\section{Results and discussion}
\subsection*{Gene--centric comparative approach}
For each Drosophila species examined (listed in Tab.1 and referenced to as D.xxx), we compile a list of genes orthologous to a D.melanogaster (D.mel) gene, according to the ``12 drosophila genomes project" data (Tab.1 and Material and Methods). For each pair of D.mel/D.xxx orthologous genes, we extract in both species the upstream, downstream and intronic regions. Upstream and downstream regions are extracted up to the next neighboring gene (see Material and Methods for more details), taking the longest transcript as a reference in case of multiple transcripts. All sequences have been previously masked for repeats using the RepeatMasker program \cite{Repeat}. At this stage, the comparison procedure crucially depends on the availability of genomic annotations (i.e. gene coordinates and orthology relationships). The orthologous regions are then aligned using a local alignment procedure described later. For the alignment, the orthologous regions are oriented such that the corresponding genes are in the same orientation. Using this gene­centric approach, most intergenic regions are considered twice. For example, the region chr4:64404-68333 in D.melanogaster is first considered as the upstream region of the \textit{PlexB} gene, and then as the downstream region of the \textit{ci} gene. This redundancy is taken care of in the post-processing step, described later.

\subsection*{Alignment procedure}
For each pair of orthologous D.mel/D.xxx genes, we respectively align their upstream regions, downstream regions and introns. This is done by orienting the transcripts in the same direction, such as to distinguish same from opposite strand. Local pairwise alignments between orthologous sequences was performed using CHAOS \cite{brudno}, which is an heuristic alignment algorithm with some peculiar features optimized for large non coding DNA sequences. CHAOS works by chaining small words (called seeds) that match between the two input sequences. Unlike BLAST, it is a double seed technique which allows some degeneracy in seeds. It chains toghether seeds that are closer than a maximum distance d and it returns the highest scoring chains, according to a standard Needleman­Wunsch metric. These highest scoring chains constitue the conserved non­coding blocks (CNBs). Because it is a local alignment, it is able to identify non­syntenic CNBs order with a very high resolution. Moreover, because the alignment is performed on both strands, we also identify CNBs resulting from inversion events. Also, it is able to rapidly align large sequences with a better specificity than purely local aligners, thanks to the double seed technique. We choose a quite sensitive set of parameters in CHAOS (see Material and Methods). An assessment of statistical significance of alignment scores is introduced to discriminate true from random alignments. The scoring cutoff is calculated by aligning randomly selected non-orthologous sequences, and setting a false discovery rate (FDR) of $2\cdot10^{-3}$ .

\subsection*{Matherials and methods}
Sequences and annotations have been downloaded from the AAA site \cite{aaa} as fasta and GFF3 files respectively. The Drosophila melanogaster sequences and annotations correspond to version 4.3. For the other drosophila, the sequences correspond to the CAF1 assemblies and are now available from GenBank. The annotations result from a reconciliation procedure of various annotations, whereas the homology maps are built using a fuzzy reciprocal blast. For details, see \cite{aaa}.\\

We rely on the gene annotations to extract the D.melanogaster introns. When several transcripts exist for a single gene, we consider the longest transcript and its introns. For other drosophila species, no annotations exist for intron/exon structure. Hence, we extract the locus corresponding to the full gene, and align it using CHAOS to each intron of the orthologous gene. For intergenic regions, we applied a conservative definition. We define the upstream region as the longest consecutive sequence of non­exonic, non­intronic nucleotides on the 5' end of the longest transcript, and similarly for the downstream region. While this is an intuitive definition in general, it has particular implications in the case of nested genes. For a gene A nested inside the intron of a gene B, the intergenic regions associated to gene A will start at the 5'/3' extremities of gene B, in order to respect the previous definition.\\

We use CHAOS with the following set of parameters: chaos -wl 7 -co 12 -b -v -rsc 1500. The last parameter is a very loose lower threshold on the alignment score, but we apply more stringent thresholds in the post­processing step.

\subsection*{Post--processing and availability}
As mentioned previously, sequences are often considered and aligned twice, resulting in redundant CNBs. We eliminate this redundancy by scanning the output of the alignments, and merging overlapping CNBs. More precisely, we merge two CNBs if they meet all of the following requirements: (i) they overlap in D.melanogaster, (ii) they overlap in the other species, (iii) both blocks are in the same orientation in D.melanogaster, and in the same orientation in the other Drosophila species. Conditions (i) and/or (ii) are for example not fulfilled in the case of duplications; in this case, the CNBs are not merged and appear as distinct blocks. Each block is assigned a unique identifier and is labelled with its score, percentage identity, as well as with the name of the gene(s) in the surrounding of which it is located. For the reasons mentioned previously, a block often refers to its two flanking genes.\\

The full collection of CNBs for all eleven pairwise comparisons is available as a queryable database, named DrosOCB (for Drosophila Conserved Blocks). It can be accessed through a user­interface which allows to query a particular gene or a genomic region. Our database is linked with the UCSC genome browser \cite{ucsc}, such that CNBs can be displayed in their genomic context with the browser.

\subsection*{DrosOCB database content}
\begin{figure}
\begin{center}
\setlength{\unitlength}{5mm}
\includegraphics[height=6 cm ,width=8 cm]{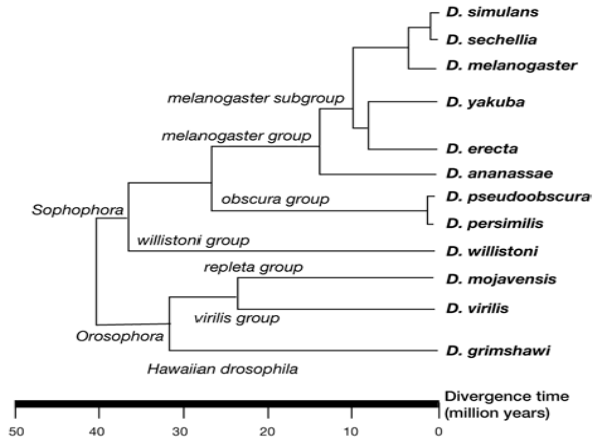}
\end{center}
\caption{The evolutionary tree of Drosophila genus, according to the ``12 drosophila genomes project". This is taken from the AAA web site [6].}
\label{fig1}
\end{figure}

\begin{figure}
\begin{center}
\setlength{\unitlength}{5mm}
\includegraphics[height=6 cm ,width=12 cm]{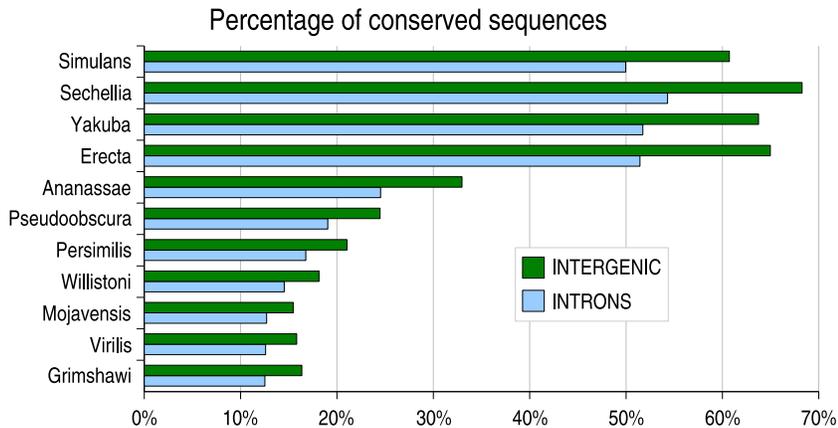}
\end{center}
\caption{Percentage of conserved sequences in intergenic and intronic regions for each of the 11 species compared with D.melanogaster The percentages are determined taking the total length of the intergenic/intronic CNBs (redundant CNB portions are counted only once), and dividing by the total non­redundant length of the aligned intergenic/intronic sequences. This correponds to columns 5 and 6 of Table 1.}
\label{fig2}
\end{figure}

In Table 1, the content of the database is summarized for each species compared with D.melanogaster. Their phylogenetic relationship is shown in Fig. \ref{fig1}. The cumulated size of the D.melanogaster sequences (intronic and intergenic) which are aligned varies in the range between 78.6 Mbp and 87.7 Mbp, depending on the total number of orthologous genes between D.melanogaster and the other species. Considering that the D.melanogaster genome size is around 120 Mbp, this means that we aligned between 65\% and 73\% of the D.melanogaster genome. Analyzing the catalog of CNBs, we can make some observations about the conservation features of Drosophila genus at large scale. The estimated percentage of non coding sequences evolutionary constrained in Drosophila genome is reported in Tab. \ref{tab1} and displayed in Fig. \ref{fig2}. As expected, the percentage of conservation follows the evolutionary distance. It varies between 16\% (13\%) for D.melanogaster/D.virilis intergenic (intronic) sequences, the most evolutionary distant species in the phylogenetic tree, and 68\% (54\%) for D.melanogaster/D.sechellia. These estimations are lower than the ones obtained for D.melanogaster compared with Drosophila D.virilis, D.pseudoobscura and D.yakuba from previous work [9­11]. However, we applied a rather conservative threshold on the scores of the CNBs, such as to reduce the number of spurious alignments. These conservation percentages are always higher in intergenic regions as compared to intronic regions. However, these figures should be taken with some care, as some regions, labelled as ``intergenic" in some drosophila species (and thus not aligned as introns) might well turn out to be intronic, as distant exons will become better annotated. In fact, whereas the mean size of genes in D.melanogaster is 6.1 kb, it ranges from 2.9 kb (D.sechellia) to 4.1 kb (D.virilis) for the other species, indicating that some gene annotations might still miss distant exons.\\

Interestingly, these proportions are roughly constant inside the melanogaster subgroup (around 50\%), indicating that the difference in the evolutionary distance between D.melanogaster and D.simulans/D.sechellia on one hand (about 5 My), and D.yakuba/D.erecta (about 10 My) on the other hand is too small to affect the conservation of non­coding DNA. There is a important decrease outside this branch (roughly 25\% for D.ananassae, the closest species outside the melanogaster subgroup). The percentages for species outside the Sophophora subgenus (D.virilis, D.mojavensis and D.grimshawi) are again very comparable (about 15\%). The mean size of CNBs obtained in our output is comprized between 50 bp and 94 bp, and increases with decreasing evolutionary distance, as expected (cf. column 6 in Tab.1). It is shorter for intronic regions than for intergenic regions. The lower threshold of 50 bp indicates that, although the alignment procedure is sensitive in allowing some degeneracy in the compared sequences, it preserves a certain degree of selectivity, discarding very short isolated CNBs with a score below the cutoff threshold.\\

\begin{figure}
\begin{center}
\setlength{\unitlength}{5mm}
\includegraphics[height=8 cm ,width=14 cm]{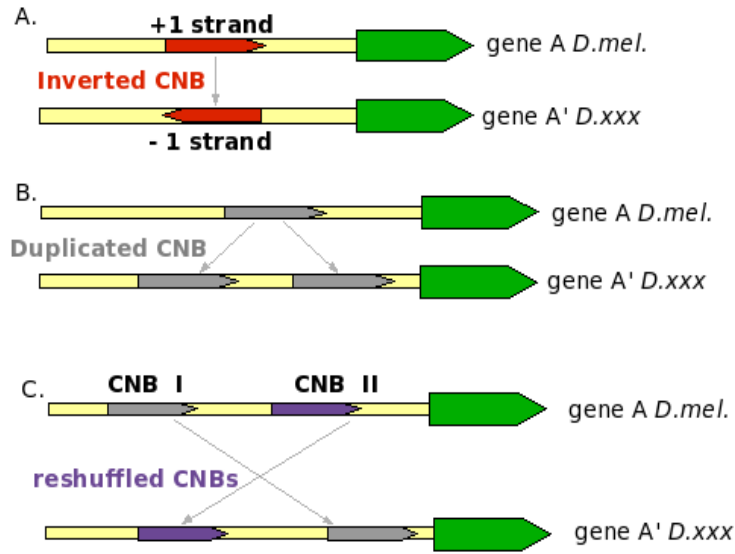}
\end{center}
\caption{Rearrangements events of CNBs The picture represents typical genomic rearrangements that can be observed in DrosOCB. In the case A, the CNB (red box in the picture) is a conserved sequence that has changed is orientation in one of the two species, taking as common reference the orientation of the transcript. Case B represents a CNB which is duplicated in the second species compared with D.melanogaster. In this case, the same D.melanogaster sequence matches two different regions in the compared species, and appears as two overlapping blocks. Case C shows a third case of genomic rearrangement that can be detected in the DrosOCB content. We called "reshuffled CNBs" two sequences that are conserved in non collinear order in the two species (purple box in the picture). Note that we have not depicted here the most frequent configuration of non­inverted, collinear CNBs.}
\label{fig3}
\end{figure}

\begin{figure}
\begin{center}
\setlength{\unitlength}{5mm}
\includegraphics[height=5 cm ,width=10 cm]{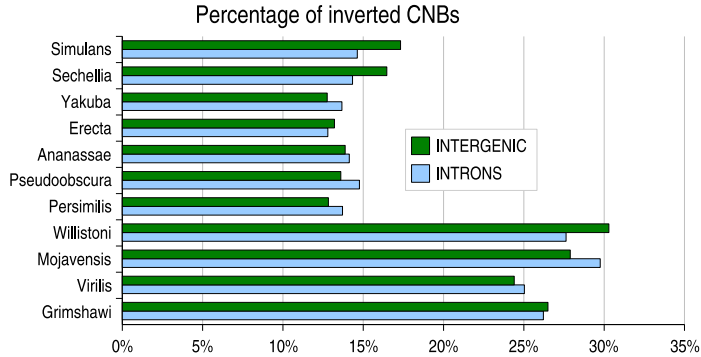}
\end{center}
\caption{Percentage of inverted CNBs in intergenic and intronic sequences in DrosOCB. These percentages are computed by taking the ratio of the number of inverted CNBs divided by the total number of blocks, in intergenic and intronic regions respectively. Note that the length of the blocks is not taken into account here.
}
\label{fig4}
\end{figure}

\begin{figure}
\begin{center}
\setlength{\unitlength}{5mm}
\includegraphics[height=5 cm ,width=10 cm]{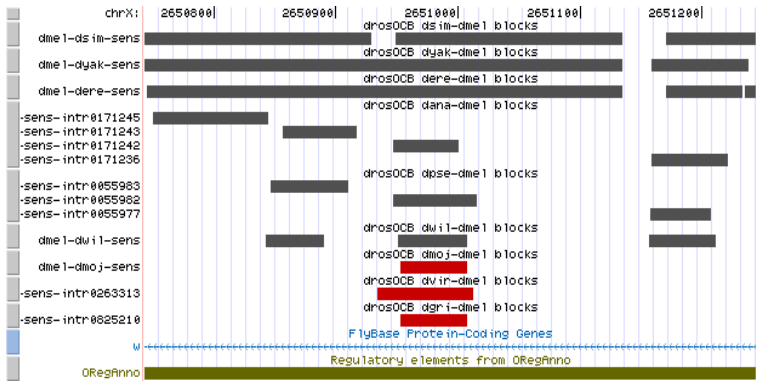}
\end{center}
\caption{Example of inverted CNBs UCSC genome browser window with our custom tracks, showing an example of specific lineage inversion event in Melanogaster region X:5,494,877-5,495,622. Grey colour coded boxes represents DrosOCB CNBs conserved across all species in the same orientation respect to the Melanogaster locus. The red boxes
(tracks dmoj-dmel, dvir-dmel, dgri-dmel) represent inverted CNBs in all species with respect to Melanogaster, highlithing an inversion event specific of the branch common to D.mojavensis, D.virilis and D.grimshawi (see also Figure 1).
}
\label{fig5}
\end{figure}

Due to the fact that we use a sensitive local alignment procedure, we are able to spot small scale genomic rearrangements that are not visible in standard alignments (see Fig.  \ref{fig3}). As an illustration, we will focus on a particular feature, namely inverted CNBs. By this, we mean CNBs that lie on opposite strands in the orthologous regions, a situation which might result from local genomic inversions in one of the two species. Since the drosophila genomes are known to have an extreme plasticity at large/medium scales \cite{ranz}, it is interesting to verify whether this is also true in or below the kb range. Fig. \ref{fig4} plots the percentage of CNBs that are inverted, for all eleven pairwise alignments, for intronic and intergenic regions. Depending on the evolutionary distance, this percentage ranges from 15\% to almost 30\%. Interestingly, these percentages are very comparable for intergenic and intronic regions, indicating that the evolutionary dynamics is similar for these regions \cite{bergman}. In the custom track provided for UCSC genome browser, we use a particular color coding to distinguish between CNBs on the same strand (grey boxes) and the inverted CNBs (red boxes). Figure \ref{fig5} shows an interesting example of such an event, in one of the introns of the white gene on chromosome X. The central region is highly conserved in all pairwise alignments, but the CNBs are inverted in the pairwise alignments with D.mojavensis, D.virilis and D.grimshawi. This is not due to an inversion of the full gene locus, since the transcripts are all taken in the same orientation when performing the alignment. The inverted CNBs have a very high score and a conservation of 90\% over 66 bp, excluding that they might be spurious alignments. Hence, one can speculate that there exists a local inversion inside this intron which appeared in the common branch of these three Drosophila species. Note that we applied a high threshold on the score of the DrosOCB blocks (2200, $FDR=0.8\cdot10^{-3}$ ) such as to reduce the number of displayed blocks. Interestingly, the ORegAnno track \cite{Montgomery} in the lower part of the UCSC window indicates the presence of a regulatory element, more precisely an enhancer, underlying that even functional elements are subject to extensive rearrangements, as previously noted \cite{ludwig,eisen,stupka}.

\section{Conclusions}
In this paper, we have described a new, local but genome­wide alignment procedure for binary comparisons of Drosophila melanogaster with eleven currently available drosophila genomes. We have shown that the resulting collection of CNBs, organized in the DrosOCB database, constitutes a high­resolution collection of non­coding DNA conservation in drosophila. The small size of the blocks, and the local nature of the alignment highlights small scale genomic rearragement events, that are not apparent from other approaches. As a preliminary study, we have focused on inverted CNBs which might correspond to small scale inversions. A more detailed analysis of this phenomenon and other rearragements is left for a future more complete investigation. An interesting aspect of our preliminary analysis is that the localization of these inverted blocks is not evenly distributed among chromosomes. Chromosome X seems to have a much higher than average proportion of these inverted blocks, indicating that it has undergone more extensive rearragements than the autosomal chromosomes, as noted previously \cite{Gonzales}. The alignment procedure described in this work provides an optimal tool for a high resolution comparison of non coding DNA sequences. The content of the database, and the observed high rate of low scale reshuffling suggest that this database of CNBs can constitute the starting point for several investigations, related to the evolution of regulatory DNA in Drosophila, the in silico identification of unannotated functional elements and the search for transcription factor binding sites.

\section*{Acknowledgements}
\textit{This work was partially supported by FIRB grant RBNE03B8KK from the Italian Ministry for
Education, University and Research. The authors would like to thank Pierre Mouren and Davide Cor\`a for useful suggestions and technical support.
}

\begin{sidewaystable}[htbp]
\centering
\begin{tabular}{|l|c|c||c|c||c|c||c|c||}
\hline
&\textbf{Genome }&\textbf{Number}&\multicolumn{2}{c||}{\textbf{Size of }}&\multicolumn{2}{c||}{\textbf{\% of }}&\multicolumn{2}{c||}{\textbf{mean size}}\\

&\textbf{size*}&\textbf{orthologous}&\multicolumn{2}{c||}{\textbf{aligned}}&\multicolumn{2}{c||}{\textbf{conserved}}&\multicolumn{2}{c||}{\textbf{of CNBs}}\\

&\textbf{(Mbps)}&\textbf{genes}&\multicolumn{2}{c||}{\textbf{sequences (Mbps)}}&\multicolumn{2}{c||}{\textbf{sequences}}&\multicolumn{2}{c||}{\textbf{(bps)}}\\
\hline
&&&intronic&intergenic&intronic&intergenic&intronic&intergenic\\
\hline
D.sim&139,8&11540&36,9&55,4&50\%&61\%&103&118\\
\hline
D.sec&168,9&12074 & 39,2 & 56,7&54\%&68\%&102&115\\
\hline
D.yak&168,0&13005&45,7&62,5 &52\%&64\%&84&94\\ 
\hline
D.ere&154,9&12459& 44,1 & 60,7&51\%&65\%&88&95\\
\hline
D.ana&234,3&11530&47,7 &61,7 &25\%&33\%&57&59\\
\hline
D.pse&154,9&11795&66,9 &44,1& 19\%&24\%&52&54\\
\hline
D.per&191,0&10657&39,4 & 58,3& 17\%&21\%&53&54\\
\hline
D.wil&238,9&10907& 52,5&79,9 &15\%&18\%&49&49\\
\hline
D.moj&196,6&11213&51,1&76,6 &13\%&15\%&49&49\\
\hline
D.vir&209,0&11346& 51,6 & 77,2 &13\%&16\%&49&49\\
\hline
D.gri&203,3&11541 &48,9 &71,7 &13\%&16\%&49&49\\
\hline
\end{tabular}
\caption{DrosOCB database content summary From the left, columns indicate the Drosophila species, the size of the genome (defined as the total size of the genome fasta files downloaded), the number of orthologous genes between D.melanogaster and the second species, the size of intronic and intergenic sequences aligned, the percentage of conserved sequences (i.e. the total non­redundant size of intronic and intergenic CNBs divided by the size of intronic/intergenic sequences aligned), the mean size of intronic and intergenic CNBs.}
\label{tab1}
\end{sidewaystable}

\end{document}